\documentclass[aps,prl,twocolumn,showpacs,floatfix,preprintnumbers,amsmath,amssymb]{revtex4}

\usepackage{epsfig}
\usepackage{color}
\usepackage{url}
\usepackage{graphicx}

\begin{document}

\preprint{SLAC-PUB-12934}
                                                                                                                                                          
\title{ A Simple Explanation for the $X(3872)$ Mass Shift Observed for Decay to $D^{*0} \bar{D^0}$}

 \author{W. Dunwoodie}
\affiliation{    Stanford Linear Accelerator Center, Stanford, California 94309, USA}
 \author{V. Ziegler}
\affiliation{    Stanford Linear Accelerator Center, Stanford, California 94309, USA}

\date{\today}

\begin{abstract}
  We propose a simple explanation for the increase of approximately
 3 MeV/c$^2$  in the mass value of the $X(3872)$ obtained from
 $D^{*0} \bar{D^0}$ decay relative to that obtained from decay to $J/\psi \pi^+ \pi^-$.
 If the total width of the $X(3872)$ is 2-3 MeV, the peak position in the
 $D^{*0} \bar{D^0}$ invariant mass distribution is sensitive to the final
 state orbital angular momentum because of the proximity of the $X(3872)$
 to $D^{*0} \bar{D^0}$ threshold. We show that for total width 3 MeV and one
 unit of orbital angular momentum, a mass shift $\sim$3 MeV/c$^2$ is
 obtained; experimental mass resolution should slightly increase this
 value. A consequence is that spin-parity $2^-$ is favored for the
 $X(3872)$.

\end{abstract}

\pacs{14.40.Gx, 13.25.Hw}

\keywords{$X(3872)$}

\maketitle

  The $X(3872)$ has been seen primarily in its $J/\psi \pi^+ \pi^-$ decay
 mode~\cite{1a,1b,1c,1d,1e}, from which a measured mass value of $3871.4\pm 0.6$ MeV/c$^2$
 is obtained~\cite{2}; an upper limit on the total width of 2-3 MeV has been
 estimated~\cite{1a}. Observation of decay to the $J/\psi \gamma$ final state~\cite{3a,3b}
 has established positive C-parity, and analysis of the decay
 angular distributions~\cite{4} has narrowed the spin-parity ($J^P$)
 possibilities to $1^+$ or $2^-$ for the $X(3872)$. The invariant mass
 distribution for the $D^0 \bar{D^0} \pi^0$ system resulting from B-decay to
 the $D^0 \bar{D^0} \pi^0 K$ final state shows a peak near threshold yielding
 a mass value $3875.2\pm 0.7 +0.9 -1.8$ MeV/c$^2$~\cite{5}, and this has been
 interpreted as evidence for the decay process $X(3872)\rightarrow D^{*0} \bar{D^0}$. A
 recent BaBar analysis~\cite{6} has confirmed this result, and has obtained
 the corresponding mass value $3875.1^{+0.7}_{-0.5}\pm 0.5$ MeV/c$^2$. In each
 case, the first error quoted is statistical, and the second is
 systematic. These results are in excellent agreement.
 Compared to the mass value $3871.4\pm 0.6$ MeV/c$^2$~\cite{2} for 
 the $J/\psi \pi^+ \pi^-$ decay mode, the mass difference is $3.8^{+1.2}_{-2.0}$ MeV/c$^2$ 
 from Belle, and $3.7^{+1.1}_{-0.9}$ MeV/c$^2$ from BaBar.  The significance of the latter
 is at the four standard deviation level, and given the
 consistency of the BELLE and BaBar results for $D^{*0} \bar{D^0}$, it would
 seem to be a real effect. We take the point of view that this is
 indeed the case, and suggest a possible explanation, which, although
 very simple, carries some significant physical implications.
 
  Consider the decay process $B\rightarrow K X$, with $X$ is a resonance decaying
 to a final state $F$. The invariant mass distribution for the system
 $F$ takes the form
\begin{eqnarray} 
 \frac{{\rm d} N}{{\rm d} m} = C_1 m \frac{ |T_B(m)|^2 (p/m_B)\int{ |T_F(m)|^2  {\rm d} \phi_{\it F}(m) }}{\left( m_0^2 - m^2\right )^2 + m_0^2  \Gamma_{tot}(m)^2},
\end{eqnarray} 
\noindent
 where  $C_1$ is a constant,
 \noindent
          $m$ is the invariant mass of system $F$,
 \noindent
     $T_B(m)$ is the invariant amplitude describing the $B$ to $K X$ coupling,
 \noindent
     $T_F(m)$ is the invariant amplitude describing the $X$ to $F$   coupling,
 \noindent
  ${\rm d}\phi_{\it F}(m)$ is the element of  $F$ decay phase space,
 \noindent
 and the denominator is the square of the relativistic Breit-Wigner
 propagator describing the resonance $X$. The factor m is present because
 the Lorentz-invariant phase space volume element is proportional to
 ${\rm d}m^2$, and $(p/m_B)$ is the two-body phase space factor for $B\rightarrow K X$ decay,
 with
 \begin{eqnarray}
     p = \frac{\sqrt{\left [m_B^2 - (m_K + m)^2\right ]\left [m_B^2 - (m_K - m)^2\right ]}}{2 m_B}.
  \end{eqnarray}
 
 Equation (1) can be written in terms of the partial width for
 $X$ decay to $F$, $\Gamma_F(m)$, as

 \begin{eqnarray}
 \frac{{\rm d}N}{{\rm d}m}=\frac{C_2\cdot  m\cdot  p\cdot  |T_B(m)|^2  \Gamma_F(m)}{\left (m_0^2 - m^2\right )^2 + m_0^2  \Gamma_{tot}(m)^2}.
\end{eqnarray}
 
 In general,  $\Gamma_{tot}(m)$, which is the mass-dependent total width
 of $X$, takes the form

  \begin{eqnarray}
    \Gamma_{tot}(m) = \sum_{i=1}^{M} \Gamma_i(m),
 \end{eqnarray}
 where $M$ is the number of decay modes of $X$, and the $\Gamma_i$ are the
 individual partial widths, of which $\Gamma_F$ is one.

  The amplitude $T_B$ is not known. However, if $X$ has spin $J$, angular
 momentum conservation requires that there be $J$ units of orbital
 angular momentum associated with the $K X$ system resulting from $B$ decay,
 and so we express $T_B$ in terms of the corresponding centrifugal barrier
 factor as follows:
  \begin{eqnarray}
    T_B(m) \sim  \frac{ p^J}{\sqrt{ D_J(p,R) }},                                 
  \end{eqnarray}
 where $D_J(p,R)$ is the Blatt-Weisskopf Damping Factor~\cite{blatt}, and $R$ is
 the associated radius parameter, for which we choose the value
 5 GeV$^{-1}$ (i.e. 1 Fermi). The $D_J$ functions for $J=0-3$ are summarized
 in Table I. For the $X(3872)$, the mass range is limited
 (3.87-3.91 GeV/c$^2$) and the Q-value for $B\rightarrow K X$ decay is large (~0.9
 GeV/c$^2$), so that the m-dependence introduced by this description of
 the B decay vertex is small.
 
\begin{table}[ht]
\begin{center}
\begin{tabular}{l c l}
\hline 
$J$ & & $D_J(p, R)$ \\
\hline 
0 & & 1 \\
1 & & $1+(p R)^2$ \\
2 & & $9+3(p R)^2+(p R)^4$ \\
3 & & $225+45(p R)^2+6(p R)^4+(p R)^6$ \\
\hline 
\end{tabular}
\end{center}
\caption{The Blatt-Weisskopf Damping Factors.}
\end{table}

  For the $X(3872)$ only three decay modes have been observed to date~\cite{1a,1b,1c,1d,1e,3a,3b,5,6}, 
 and of these the $J/\psi \pi^+ \pi^-$ and $J/\psi \gamma$ modes are
 well above their respective mass thresholds. It follows that the
 $m$-dependence for each of these contributions to Eq.(4) is very weak.
 In contrast, for the $D^{*0} \bar{D^0}$ mode, the dependence on $m$ is potentially
 quite strong because of the proximity of the invariant mass threshold
 to the $X(3872)$ mass; this is, in fact, the main point of this Letter.
 If the state $F$ of Eq.(3) is chosen to be the $D^{*0} \bar{D^0}$ final state,
 the numerator is very sensitive to its dependence on $m$, as will be
 discussed in detail below. However, in the denominator, this mass
 dependence is subsumed into that of the total width by way of Eq.(4).
 Since the other known modes have very little mass-dependence, and
 since other as yet unobserved decay modes may contribute to $\Gamma_{tot}$,
 we choose to treat $\Gamma_{tot}$ as a constant in the computations to
 follow.
 
 From Eqs.(1) and (3),
  \begin{eqnarray}
    \Gamma_F(m) \sim \int{ |T_F(m)|^2  {\rm d}\phi_{\it F}(m) } ,              
 \end{eqnarray}
 \noindent
 and with $F$ the $D^{*0} \bar{D^0}$ final state,
  \begin{eqnarray}
    {\rm d}\phi_{\it F}(m) \sim (q/m) {\rm d}\Omega                                        
  \end{eqnarray}
 \noindent
 where
\begin{eqnarray}
 q = \frac{\sqrt{\left [m^2 - (m_D + m_D^*)^2\right ]\left [m^2 - (m_D - m_D^*)^2\right ]}}{2m}    
 \end{eqnarray}
 \noindent
 is the momentum in the $D^{*0} \bar{D^0}$ rest frame; we ignore the width of
 the $D^{*0}$ ($<$ 1 MeV~\cite{2}). Integrating over $\Omega$, we obtain
 \begin{eqnarray}
    \Gamma_F(m) \sim (q/m)  |T_F(m)|^2 ,                               
 \end{eqnarray}
 \noindent
 and following Eq.(5) (with the same value of $R$) we express $T_F(m)$ as
 \begin{eqnarray}
    T_F(m) \sim \frac{q^L}{\sqrt{ D_L(q,R) }},                                
 \end{eqnarray}
 \noindent
 where $L$ is the number of units of orbital angular momentum associated
 with the $D^{*0} \bar{D^0}$ system resulting from decay of the $X(3872)$.
 
  The full expression for the $D^{*0} \bar{D^0}$ invariant mass projection is
 then
 \begin{eqnarray}
\frac{{\rm d}N}{{\rm d} m} =C_3\frac{\left ( p^{2J+1}/D_J(p,R)\right )\left ( q^{2L+1}/D_L(q,R)\right ) }{\left (m_0^2 - m^2\right )^2 + m_0^2  \Gamma_{tot}^2}
 \end{eqnarray}
 \noindent
 where  $C_3$ is a constant.
 
 If the $X(3872)$ has $J^P=1^+$ $L$ can be 0 or 2, while for $J^P=2^-$ $L$ can
 be 1 or 3, since parity should be conserved in the decay process.
 Consequently we consider that only L values in the range 0-3 are
 of relevance to the $X(3872)$. Also we use the following central mass
 values~\cite{2} in our calculations:
  \begin{eqnarray*}
       m(D^0) &=& 1864.84 \,[\pm 0.18 ]\,\, {\rm MeV/c}^2 \\
      m(D^{*0}) &=& 2006.96 \,[\pm 0.19 ]\,\, {\rm MeV/c}^2 .\\
 \end{eqnarray*}

\noindent
  These yield the threshold mass value $3871.80\pm 0.37$ MeV/c$^2$ for
  decay to $D^{*0} \bar{D^0}$ (again ignoring the width of the $D^{*0}$), 
  where the error is obtained by combining the error on twice the $D^{0}$ mass 
  and that on the $D^{*0} - D^0$ mass difference in quadrature. 
 
  We choose $X(3872)$ mass values 3870.8, 3871.4 and 3872.0 MeV/c$^2$
  (i.e. the PDG 2007 average and plus or minus one sigma), and for each
  use $\Gamma_{tot}$ values 2, 3 and 4 MeV. For each of these nine
  combinations we use Eq.(11) to compute the lineshape for the choices
  $L=0$ and 2 with $J=1$ (i.e. $X(3872)$ $J^P = 1^+$), and $L=1$ and 3 with
  $J=2$ (i.e. $X(3872)$ $J^P = 2^-$). For $L=2$, all of the lineshapes obtained
  are very broad, reaching a maximum close to mass 3.90 GeV/c$^2$, while
  for $L=3$ the distribution obtained increases monotonically through
  3.91 GeV/c$^2$, the upper limit of the region investigated. It follows
  that for $L=2$ and $L=3$ Eq.(11) yields behavior which is totally
  unlike that observed for data~\cite{5,6}, and so the possibility that
  such contributions play a significant role in $X(3872)$ decay to
  $D^{*0} \bar{D^0}$ is discarded.
 
   For the remaining ($J^P=1^+: L=0, J=1$) and ($J^P=2^-: L=1, J=2$)
  possibilities, Eq.(11) does lead to lineshapes which peak at mass
  values a few MeV/c$^2$ larger than the input $X(3872)$ values.  The
  mass shift is defined as the difference between the observed peak 
  mass position and the input $X(3872)$ mass value, $m_0$, and the 
  mass shifts obtained are summarized in Tables II and
  III respectively. In both Tables, the mass shift increases with
  decreasing $m_0$ and increasing $\Gamma_{tot}$. However each
  value in Table II is smaller than the corresponding one in Table III,
  the difference averaged over the Tables being $\sim$1.85 MeV/c$^2$.
  Furthermore, it seems from Table II to be unlikely
  that a mass shift of $\sim$3 MeV/c$^2$ can be obtained for a reasonable
  choice of $X(3872)$ mass and width, especially since the $D^{*0} \bar{D^0}$
  mass distribution which results becomes considerably broader as the
  $X(3872)$ mass is decreased and its width increased.

\begin{table}[ht]
\begin{center}
\begin{tabular}{c|c c c| c c c |c c c}
\hline
  &\multicolumn{9}{c}{$\Gamma_{tot}$ [MeV]}\\
\cline{2-10}
$X(3872)$ mass [MeV/c$^2$] & & 2 & & & 3 & & & 4 & \\
\hline
  3870.8 & & 1.54 &  & & 1.75 &  & & 1.99 &   \\  
  3871.4 & & 0.90 &  & & 1.17 &  & & 1.45 &  \\  
  3872.0 & & 0.45 &  & & 0.75 &  & & 1.02 &  \\  
\hline
\end{tabular}
\end{center}
\caption{Dependence of the peak mass shift (in MeV/c$^2$) 
 on $X(3872)$ mass and total width for $L=0$ and $J=1$ (i.e. $J^P=1^+$).}
\end{table}

\begin{table}[ht]
\begin{center}
\begin{tabular}{c|c c c| c c c |c c c}
\hline
  &\multicolumn{9}{c}{$\Gamma_{tot}$ [MeV]}\\
\cline{2-10}
$X(3872)$ mass [MeV/c$^2$] & & 2 & & & 3 & & & 4 & \\
\hline
  3870.8 & & 3.79 &  & & 4.25 &  & & 4.75 &   \\  
  3871.4 & & 2.37 &  & & 2.99 &  & & 3.61 &  \\  
  3872.0 & & 1.28 &  & & 1.98 &  & & 2.66 &  \\  
\hline
\end{tabular}
\end{center}
\caption{Dependence of the peak mass shift (in MeV/c$^2$)
 on $X(3872)$ mass and total width for $L=1$ and $J=2$ (i.e. $J^P=2^-$).}
\end{table}
 
   Using $X(3872)$ mass 3871.4 MeV/c$^2$ and width 3 MeV, we illustrate
  the lineshape behavior obtained from Eq.(11) for $J^P=1^+$ and $J^P=2^-$
  in Fig.1 and Fig.2 respectively. In each figure, the curve is
  obtained using Eq.(11), and is then used to generate the 3000 events
  shown in the histogram below, which uses the mass intervals
  from the BaBar analysis~\cite{6}. The peak shift in Fig.2(a) agrees
  well with the result from experiment~\cite{5,6}, and the observed
  signal shapes seem better represented by that of Fig.2(b) than by
  that of Fig.1(b). However, it must be acknowledged that the
  experimental uncertainties are significant, and that 
  even the uncertainty in the location of the $D^{*0} \bar{D^0}$
  threshold (above) is relevant on the scale of the
  effect under discussion.
 
\begin{figure}[hb]
  \centering\small
  \includegraphics[width=.31\textwidth]{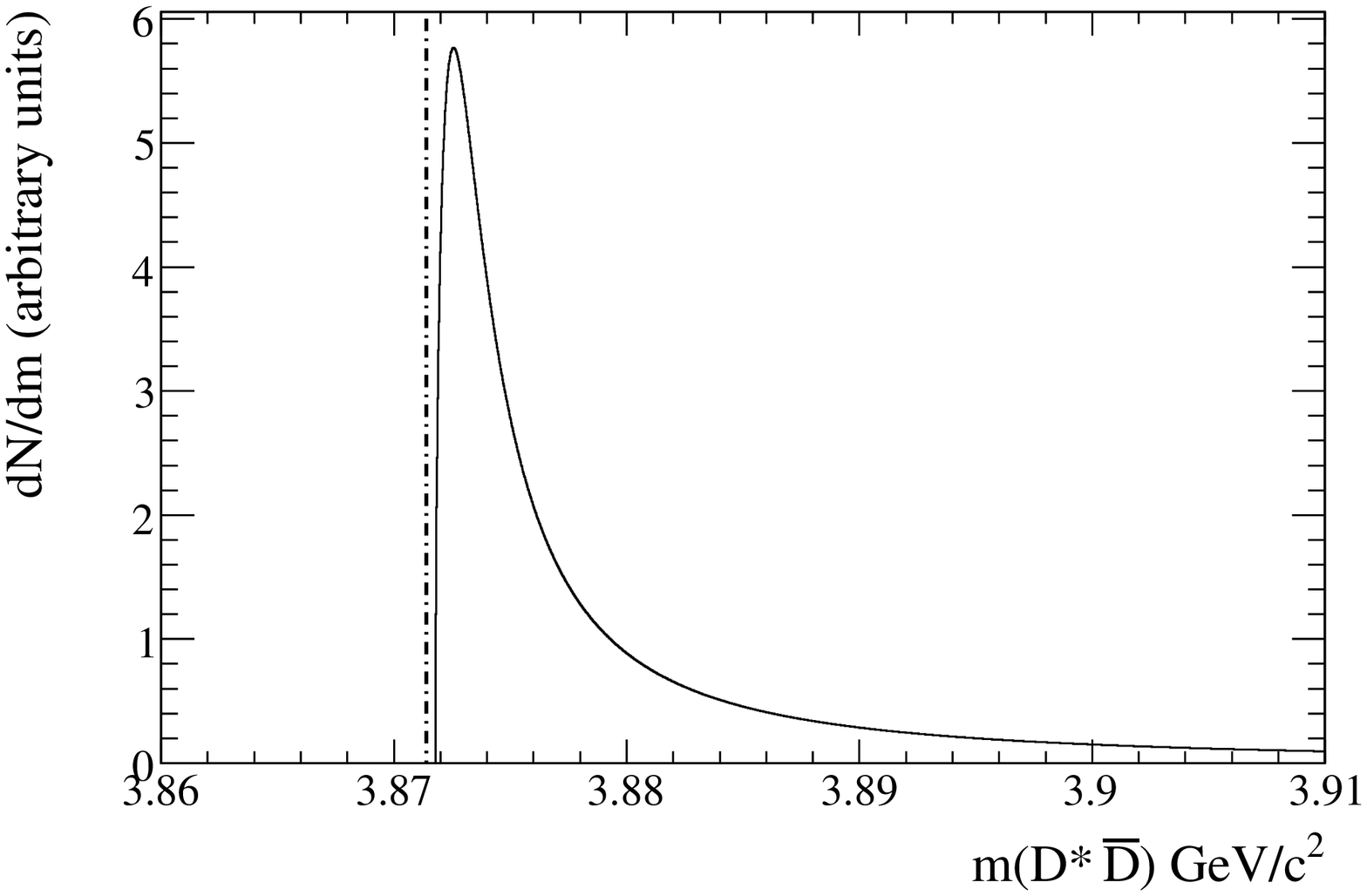}
  \includegraphics[width=.31\textwidth]{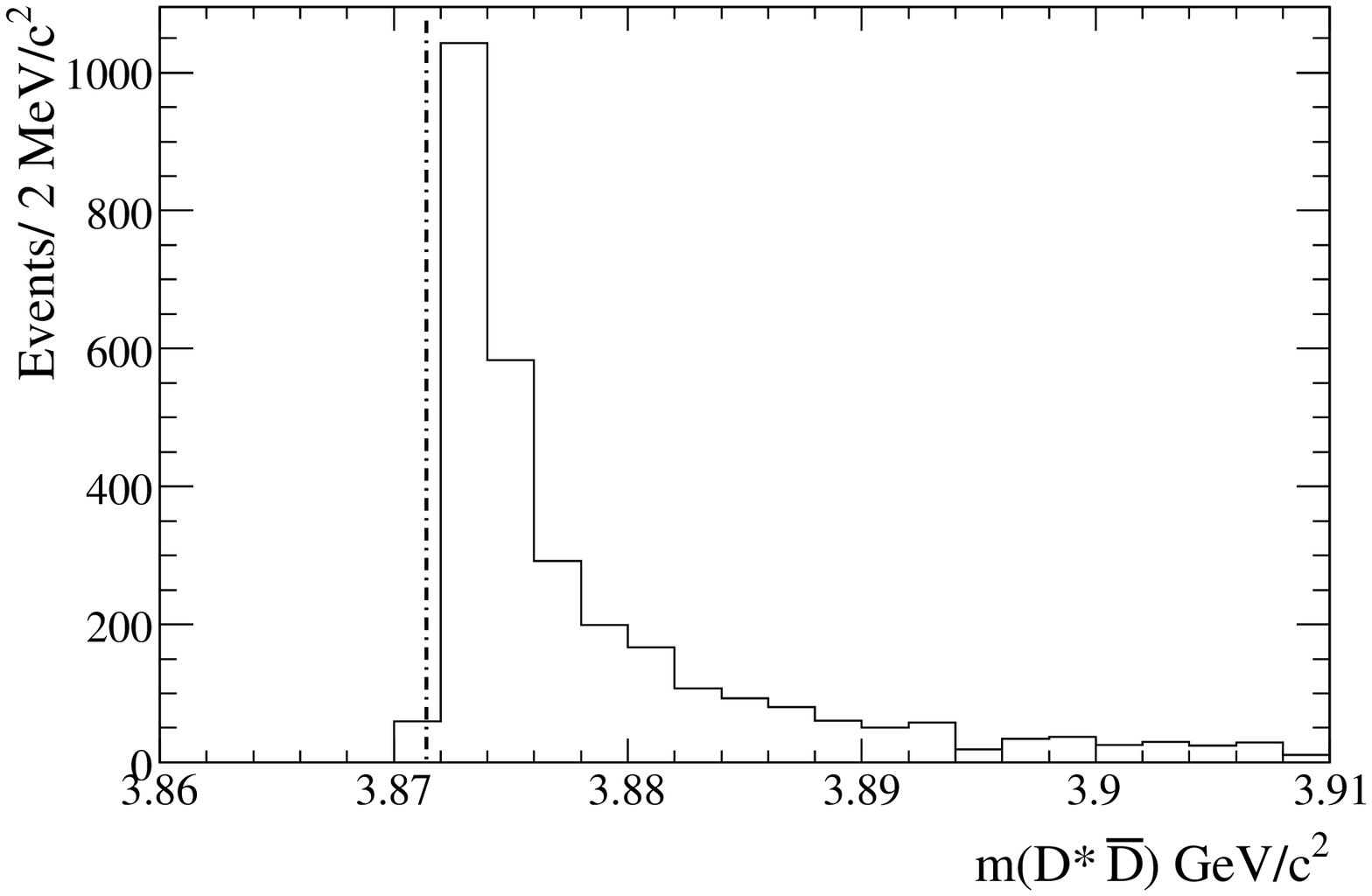}
  \begin{picture}(0.,0.)
    \put(-135,192){\bf{(a)}}
    \put(-136,85){\bf{(b)}}
    \put(-60,192){$J^P=1^+$}
    \put(-60,182){($L=0$)}
    \end{picture}
  \caption{(a) The $m(D^{*0} \bar{D^0})$ lineshape obtained from Eq.(11) for $J^P=1^+$ 
using $X(3872)$ mass 3871.4 MeV/c$^2$ (indicated by the dot-dashed line) and width 3 MeV.
(b) The histogram for 3000 events generated using the curve shown in (a).}
\end{figure} 

\begin{figure}[ht]
  \centering\small
  \includegraphics[width=.31\textwidth]{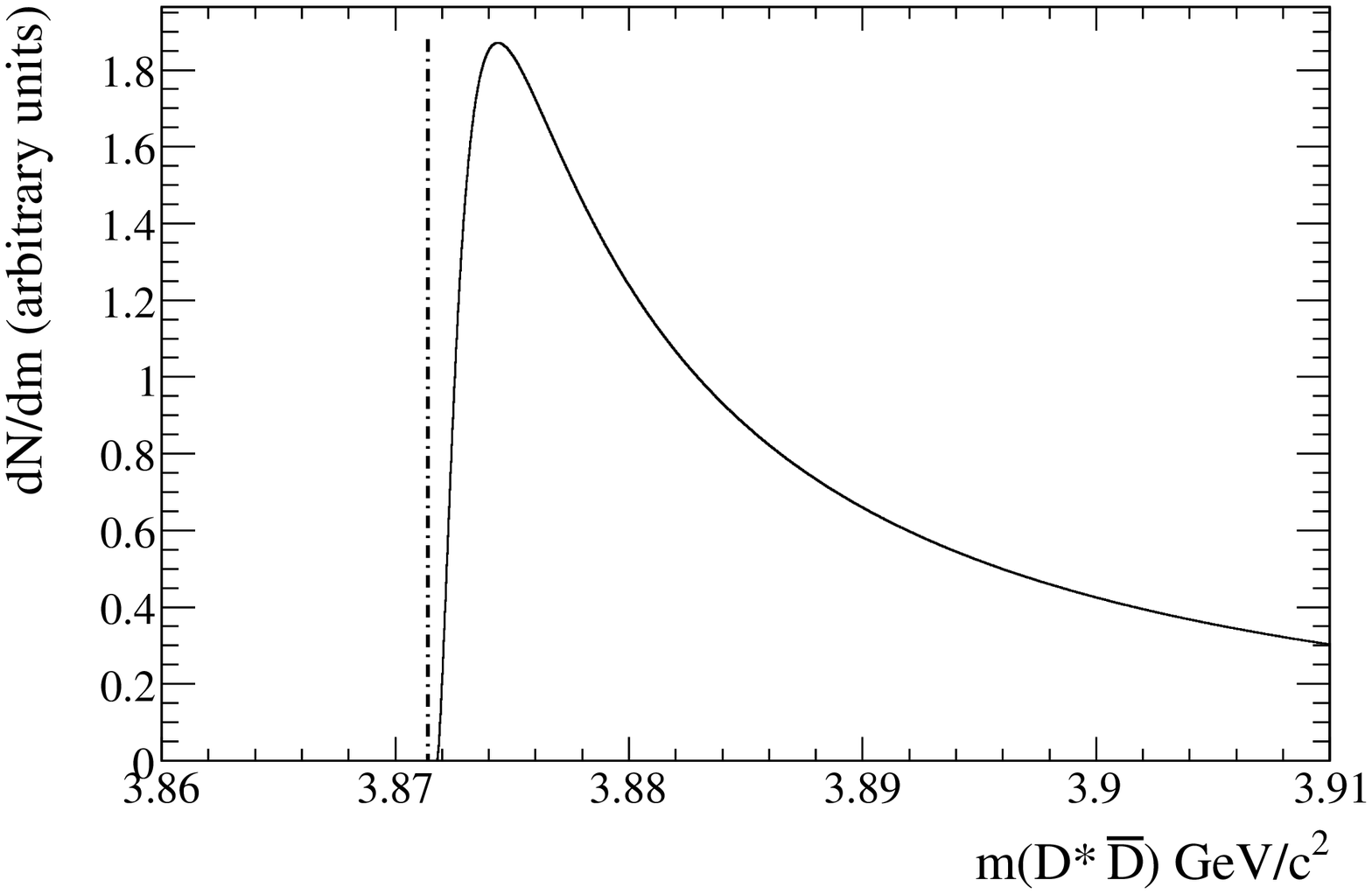}
  \includegraphics[width=.31\textwidth]{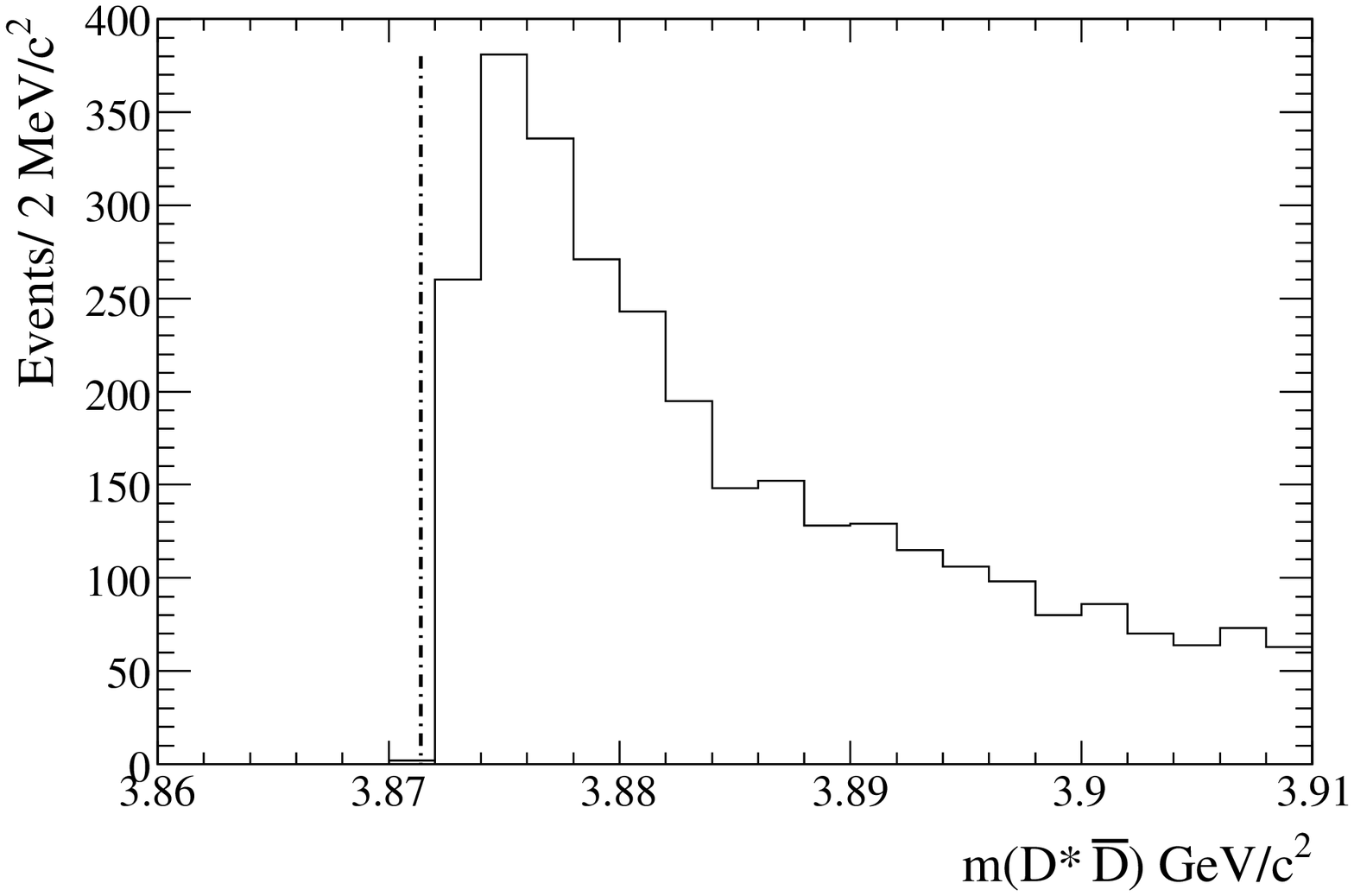}
  \begin{picture}(0.,0.)
    \put(-135,192){\bf{(a)}}
    \put(-136,85){\bf{(b)}}
    \put(-60,192){$J^P=2^-$}
    \put(-60,182){($L=1$)}
    \end{picture}
  \caption{(a) The $m(D^{*0} \bar{D^0})$ lineshape obtained from Eq.(11) for $J^P=2^-$ 
using $X(3872)$ mass 3871.4 MeV/c$^2$ (indicated by the dot-dashed line) and width 3 MeV.
(b) The histogram for 3000 events generated using the curve shown in (a).}
\end{figure}

  We have made no attempt to study the effect of detector resolution
  on the mass shifts calculated above. Near threshold, such effects
  should not be represented by Gaussian smearing in mass, as is done
  usually, since this will yield contributions below threshold. It is
  three-momentum resolution which is the source of the smearing, and
  this must be investigated by full detector simulation for the
  experiment in question. Since such simulation obviously cannot yield
  events below threshold, it seems probable that the peak mass
  shifts calculated above will be increased as a result of experimental
  resolution. We suspect that such effects will be small ($<$ 1 MeV/c$^2$),
  but a thorough investigation making use of detector simulation is
  necessary.
 
   In summary, we have shown that a simple treatment of the orbital
  angular momentum involved in $X(3872)$ decay to $D^{*0} \bar{D^0}$ can account
  for the difference between the mass measured in this mode and that
  obtained from $J/\psi \pi^+ \pi^-$ decay. The results favor $J^P=2^-$ over
  $J^P=1^+$ for the $X(3872)$, but the uncertainty in the measured mass
  difference ($3.7\pm 1.2$ MeV/c$^2$), and the absence of simulated
  detector resolution effects, prevent a definite conclusion. If our
  interpretation is correct, a corollary is that the width of the
  X(3872) cannot be much smaller than $\sim$2 MeV, since otherwise significant
  displacement of the invariant mass peak for $D^{*0} \bar{D^0}$ would not
  occur.

\vspace{.1 in}

\begin{acknowledgments}
Work supported by the U.S. Department of Energy under contract number DE-AC03-76SF00515.
\end{acknowledgments}

\vspace{2.5 in}
 
\end{document}